\documentstyle[psfig,epsf,conf-X]{article}
\begin{document} 
\small
\heading{%
%
Supernova rates in Abell galaxy clusters and implications for metallicity
}
\par\medskip\noindent
\author{%
Avishay Gal-Yam$^{1}$ and Dan Maoz$^{1}$
}
\address{%
School of Physics and Astronomy and Wise Observatory, Tel Aviv University, Tel Aviv 69978, Israel
}
\begin{abstract}
Supernovae (SNe) play a critical role in the metal enrichment of the intra-cluster 
medium (ICM) in galaxy clusters. Not only are SNe the main source for 
metals, but they may also supply the energy to eject enriched gas from galaxies 
by winds. However, measurements of SN rates in galaxy clusters have not been published to date. We have initiated a program to find SNe in 163 medium-redshift ($0.06<z<0.2$) Abell clusters, using the Wise Observatory 
1m telescope. Our program has already discovered 11 spectroscopically 
confirmed SNe at $ z=0.1 - 0.24,$ and several unconfirmed SNe. We present the
main scientific goals of this project, 
and discuss a novel explanation for the centrally 
enhanced metal abundances indicated by X-ray observations of galaxy clusters,
based on the contribution of intergalactic SNe.
\end{abstract}
\section{Scientific goals}
Our main scientific goal is to derive SN rates as a function
of various parameters such as host galaxy type and cluster environment: 
position within the cluster, cluster richness and cluster vs. field. SN 
rates can then be used to determine the current and past star formation 
rates in galaxy clusters \cite{mdp}. Our measured SN rates can replace 
the assumed rates used so far in studies of metal abundances in the 
intracluster gas.
We also intend to study the rate, distribution and properties of intergalactic SNe in galaxy clusters. We have already discovered one candidate, 
SN 1998fc (See fig. 1).\\
Our search is also sensitive to other optical transients, such as AGNs in the
clusters and behind them, flares from tidal disruption of stars by dormant
massive black holes in galactic nuclei and GRB afterglows. We may also detect
the gravitational lensing effect of the clusters on background SNe.        
\section{Intergalactic SNe and metal abundances in clusters}
The existence of a diffuse population of intergalactic stars is supported
by a growing body of observational evidence such as intergalactic planetary nebulae in the Fornax and Virgo clusters \cite{arn}, and intergalactic red giant stars in Virgo \cite{ftv}. Recent imaging of the Coma cluster reveals low surface brightness
emission from a diffuse population of stars \cite{gw}, the origin of which is attributed to galaxy disruption \cite{dmh}.
Since type Ia SNe are known to occur in all environments, there is no obvious reason 
to assume that such events do not happen within the intergalactic stellar population. SN 1998fc may well be such an event. The  
intergalactic stellar population is centrally distributed \cite{dub}.
Therefore, metals produced by intergalactic Ia SNe can provide an elegant
explanation for the central enhancement of metal abundances with type Ia
characteristics, recently detected in galaxy clusters \cite{dw}. 

%
%
\begin{figure}
\centerline{\vbox{
\psfig{figure=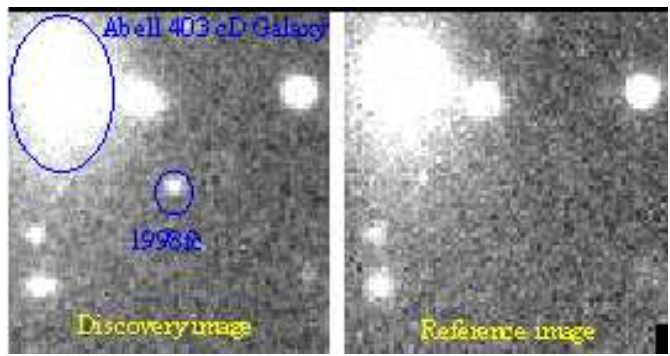,height=4.6cm}
}}
\caption[]{ The Ia SN 1998fc was detected 78 Kpc from the cD galaxy of 
Abell 403, at the cluster redshift \cite{gm2}.
This may be an intergalactic SN whose
progenitor star was a member of the diffuse intergalactic stellar
population. Alternatively, the host may be a faint dwarf galaxy. This
question could be resolved with larger number statistics of such events.
}
\end{figure}

\vfill
\end{document}